\documentclass[letterpaper]{article}

\usepackage{tikz}
\usetikzlibrary{shapes,calc,arrows,patterns,decorations.pathmorphing
	,decorations.markings}
\usetikzlibrary{arrows.meta}

\usepackage[english]{babel}
\usepackage{theorem}
\theoremheaderfont{\itshape\bfseries}
{\theorembodyfont{\itshape}

	\newtheorem{definition}{\textbf{Definition}}
	\newtheorem{theorem}{\textbf{Theorem}}
	
	\newtheorem{remark}{\textbf{Remark}}
	\newtheorem{example}{\textbf{Example}}
	\newtheorem{problem}{\textbf{Problem}}
}

\usepackage{cite}
\usepackage{amssymb}
\usepackage{algorithmic}
\usepackage{newtxtext,newtxmath}
\usepackage{amsmath}
\usepackage{amsfonts}
\usepackage{mathrsfs}

\newcommand{\T}{^{\mbox{\tiny T}}}
\newcommand{\R}{\mathbb{R}}
\newcommand{\C}{\mathbb{C}}

\newcommand{\eps}{\varepsilon}
\let\leq\leqslant
\let\geq\geqslant

\newenvironment{proof}[1][Proof]%
{\par\noindent\textit{#1:\ }}%
{\hspace*{\fill} \rule{6pt}{6pt}}
\newenvironment{proof*}[1][Proof]%
{\par\noindent\textit{#1:\ }}{}

\DeclareMathOperator{\re}{Re}

\DeclareMathOperator{\rank}{rank}

\newenvironment{system}[1]%
{\setlength{\arraycolsep}{0.5mm}\left\{ \; \begin{array}{#1}}%
	{\end{array} \right.}
\newenvironment{system*}[1]%
{\setlength{\arraycolsep}{0.5mm} \begin{array}{#1}}%
	{\end{array}}

\begin{document}
	
	\title{The role of local bounds on neighborhoods in the network for
		scale-free state synchronization of multi-agent systems} 
	\author{Anton A. Stoorvogel, Ali Saberi, and Zhenwei Liu
		\thanks{Anton A. Stoorvogel is with Department of Electrical
			Engineering, Mathematics and Computer Science, University of
			Twente, P.O. Box 217, Enschede, The Netherlands (e-mail:
			A.A.Stoorvogel@utwente.nl)} 
		\thanks{Ali Saberi is with
			School of Electrical Engineering and Computer Science, Washington
			State University, Pullman, WA 99164, USA (e-mail: saberi@wsu.edu)}
		\thanks{Zhenwei Liu is with College of Information Science and
			Engineering, Northeastern University, Shenyang 110819,
			China (e-mail: liuzhenwei@ise.neu.edu.cn)}} 
	
	\maketitle

\begin{abstract}
	This paper provides necessary and sufficient
	conditions for the existence of solutions to the state
	synchronization problem of homogeneous multi-agent systems (MAS) via
	scale-free linear dynamic non-collaborative protocol in both
	continuous and discrete time. These conditions guarantee for which
	class of MAS, one can achieve scale-free state
	synchronization. We investigate protocol design with and without
	utilizing local bounds on the neighborhood of agents. The results shows that 
	the availability of local bounds on neighborhoods plays a key role.
\end{abstract}

\section{Introduction}

The synchronization problem for multi-agent systems (MAS) has
attracted substantial attention due to the wide potential for
applications in several areas, such as autonomous vehicles,
satellites/robots system, distributed sensor network, and smart gird
(power grid), see for instance the books
\cite{bai-arcak-wen,bullobook,kocarev-book,mesbahi-egerstedt,ren-book,%
  saberi-stoorvogel-zhang-sannuti,wu-book} and references
\cite{li-duan-chen-huang,saber-murray2,ren-beard}.

Traditionally, we used networks described by Laplacian matrices for
continuous-time systems. In early work for discrete-time systems, such
as \cite{saber-murray3,tuna2,lee-kim-shim,li-duan-chen}, the networks
were described by row-stochastic matrices. In the literature, it was
never really clarified why one would use Laplacian matrices in
continuous-time and row-stochastic matrices in discrete-time. In this
early work, researchers generally assumed the network was known. In
that case, the distinction between Laplacian and row-stochastic
matrices is not that crucial because we can easily convert one in the
other if some bounds are known for the network as outlined below.

Later, the proposed protocols in the literature for synchronization of
MAS still assumed some knowledge of the communication network such as bounds
on the spectrum of the associated Laplacian matrix and the number of
agents. In this context, knowing an upper bound for the Laplacian was
viewed as reasonable and the distinction between Laplacian and
row-stochastic matrices was therefore not viewed as crucial.

As it is pointed out in \cite{studli-2017,tegling-bamieh-sandberg,%
  tegling-bamieh-sandberg-auto-23,tegling2019scalability}, the
protocols traditionally designed for synchronization suffer from
\textbf{scale fragility}. In particular, they showed that almost all
existing protocol designs at that time failed to achieve
synchronization when the network becomes too large (unless the
protocol is adapted based on the size of the network).

In the past few years, \textbf{scale-free} linear protocol design has
been subject of research in MAS literature to deal with the existing
scale fragility in MAS \cite{liu-nojavanzedah-saberi-2022-book}.  In a
``scale-free'' design the proposed protocols are designed solely based
on the knowledge of agent models and do \textbf{not} depend on
\begin{itemize}
\item Information about the
  communication network such as the spectrum of the associated
  Laplacian matrix.
\item Knowledge about the number of
  agents. 
\end{itemize}
In the context of scale-free protocol design, the distinction between
networks described by Laplacian matrices and row-stochastic matrices
is actually crucial. One can only convert Laplacian matrices to
row-stochastic matrices if you have some information about the
network. To be more precise, each agent should have a local bound on
the weighted in-degree of the network.

This brings us to a crucial question. If this local bound is not
available for discrete-time systems (and we cannot convert the
Laplacian matrix to a row-stochastic matrix), can we then still obtain
a linear scale-free protocol? It is also interesting to investigate whether this
local bound can help us in continuous-time systems.

The results in this paper are as follows, which are essentially 
necessary conditions for scale-free synchronization of linear MAS: 
\begin{itemize}
\item For discrete-time systems without availability of local bounds
  it is impossible to obtain scale-free synchronization.
\item For discrete-time systems using local bounds
  it is possible to obtain scale-free synchronization provided the
  agents are neutrally stable.
\item For continuous-time systems without availability of local
  bounds we effectively need that agents are neutrally stable,
  minimum-phase and of relative degree $1$.
\item For continuous-time systems  using local
  bounds we only need that agents are neutrally stable.
\end{itemize}
The surprising result of this paper is that in discrete-time without
local bounds we can no longer achieve scale-free synchronization. For
continuous-time agents it is still possible without local bounds to
achieve scale-free synchronization but we need additional restrictions
on the agents namely minimum-phase and relative degree $1$.

\subsection*{Notation and background}

Given a matrix $A\in \mathbb{R}^{m\times n}$, $A\T$ and $A^*$ denote
its transpose and conjugate transpose respectively. $I$ denotes the
identity matrix and $0$ denotes the zero matrix where the dimension is
clear from the context.

To describe the information flow among the agents we associate a
{weighted graph} $\mathcal{G}$ to the communication network. The
weighted graph $\mathcal{G}$ is defined by a triple
$(\mathcal{V}, \mathcal{E}, \mathcal{A})$ where
$\mathcal{V}=\{1,\ldots, N\}$ is a node set, $\mathcal{E}$ is a set of
pairs of nodes indicating connections among nodes, and
$\mathcal{A}=[a_{ij}]\in \mathbb{R}^{N\times N}$ is the weighted
adjacency matrix with non negative elements $a_{ij}$. Each pair in
$\mathcal{E}$ is called an {edge}, where $a_{ij}>0$ denotes an
edge $(j,i)\in \mathcal{E}$ from node $j$ to node $i$ with weight
$a_{ij}$. Moreover, $a_{ij}=0$ if there is no edge from node $j$ to
node $i$. We assume there are no self-loops, i.e.\ we have
$a_{ii}=0$. A {path} from node $i_1$ to $i_k$ is a sequence of
nodes $\{i_1,\ldots, i_k\}$ such that $(i_j, i_{j+1})\in \mathcal{E}$
for $j=1,\ldots, k-1$. A {directed tree} is a subgraph (subset of
nodes and edges) in which every node has exactly one parent node
except for one node, called the {root}, which has no parent
node. A {directed spanning tree} is a subgraph which is a
directed tree containing all the nodes of the original graph. If a
directed spanning tree exists, the root has a directed path to every
other node in the tree \cite{royle-godsil}.

For a weighted graph $\mathcal{G}$, the matrix
$L=[\ell_{ij}]$ with
\[
\ell_{ij}=
\begin{system}{cl}
	\sum_{k=1}^{N} a_{ik}, & i=j,\\
	-a_{ij}, & i\neq j,
\end{system}
\]
is called the {Laplacian matrix} associated with the graph
$\mathcal{G}$. The Laplacian matrix $L$ has all its eigenvalues in the
closed right half plane and at least one eigenvalue at zero associated
with right eigenvector $\textbf{1}$ \cite{royle-godsil}. Moreover, if
the graph contains a directed spanning tree, the Laplacian matrix $L$
has a single eigenvalue at the origin and all other eigenvalues are
located in the open right-half complex plane \cite{ren-book}. 

The invariant zeros of a linear system with realization $(A,B,C,D)$
are defined as all $s\in \C$ for which
\begin{equation}\label{dhg}
	\begin{pmatrix} sI-A & -B \\ C & D \end{pmatrix}
\end{equation}
loses rank.  For a linear system with transfer matrix $G$ which is
single input and/or single output then the invariant zeros can be
easily defined as those $s\in \C$ for which $G(s)=0$.

The system is called minimum phase if all invariant zeros are in the
open left half plane (continuous-time) or in the open unit disc
(discrete-time). The system is called weakly minimum-phase if all
invariant zeros are in the closed left half plane (continuous-time) or
in the closed unit disc (discrete-time) while the invariant zeros on
the boundary are semi-simple.  For a linear system with transfer
matrix $G$ which is single input (SIMO) and/or single output (MISO), an invariant
zero $s_0$ is semi-simple if and only if
\[
\lim_{s\rightarrow s_0} \frac{1}{s-s_0} G(s)
\]
is well-defined and unequal to zero.

A single-input/single-output (SISO) linear system $(A, B, C)$, has
relative degree 1 if $CB\neq 0$. The system has relative degree $r>1$,
whenever $CA^{r-1}B\neq 0$ while $CA^{r-2}B= 0$. A
multi-input/multi-output (MIMO) linear system $(A, B, C)$, has
has uniform rank with order of infinite zeros equal to one if
\[
	\rank \begin{pmatrix} sI-A & -B \\ C & D \end{pmatrix} = \rank CB
\]
for almost all $s\in \C$. 

\section{Multi-agent systems  and local bounds on neighborhoods in the network}

Consider a multi-agent systems (MAS) consisting of $N$ identical agents: 
\begin{equation}\label{eq1}
	\begin{system*}{ccl}
		x_i^+(t) &=& Ax_i(t)+Bu_i(t),\\
		y_i(t) &=& Cx_i(t),
	\end{system*}
\end{equation}
where $x_i(t)\in\mathbb{R}^{n}$, $y_i(t)\in\mathbb{R}^p$ and
$u_i(t)\in\mathbb{R}^m$ are the state, output, and input of agent $i$,
respectively, with $i=1,\ldots, N$. In the aforementioned
presentation, for continuous-time systems, $x_i^+(t) = \dot{x}_i(t)$
with $t \in \mathbb{R}$ while for discrete-time systems,
$x_i^+(t) = x_i(t + 1)$ with $t \in \mathbb{Z}$.

The communication network provides each agent with a linear combination of
its own output relative to that of other neighboring agents. In particular, each agent $i\in\{1,\cdots,N\}$ has access to the quantity:
\begin{equation}\label{zeta0}
	\zeta_i(t) = \sum_{j=1,j\neq i}^{N}a_{ij}(y_i(t)-y_j(t)).
\end{equation}
where $a_{ij}\geq0$, $a_{ii}=0$ for $i, j \in \{1,\ldots,N\}$. The topology of the network can be
described by a graph $\mathcal{G}$ with nodes corresponding to the agents in the network and
edges given by the nonzero coefficients $a_{ij}$. In particular, $a_{ij}> 0$ implies that an
edge exists from agent $j$ to $i$. The weight of the edge equals the magnitude of $a_{ij}$.

We can express the communication in the network in
terms of the Laplacian matrix $L$ associated with this weighted graph $\mathcal{G}$. In particular,
$\zeta_i$ can be rewritten as:
\begin{equation}\label{zeta}
	\zeta_i(t) = \sum_{j=1}^{N}\ell_{ij}y_j(t).
\end{equation}
where $L=[\ell_{ij}]$ is the Laplacian matrix associated with the
communication network with
\[
\ell_{ij}=-a_{ij}\;\; (i\neq j),\qquad \ell_{ii}=\sum_{j=1}^N a_{ij}
\]
for $i,j=1,\ldots,N$ where $a_{ij}$ is the weight of the edge from
node $j$ to node $i$ if such an edge exists and $a_{ij}=0$ if such an
edge does not exist.

Note that $\ell_{ii}$ can be referred to as the local weighted
in-degree of the graph, often denoted in the literature as
$d_{\text{in}}(i)$, since we have:
\[
\ell_{ii}=d_{\text{in}}(i):=\sum_{j=1}^{N}a_{ij}
\]
For the design of protocols it has turned out to be useful to have a
bound for the local weighted in-degree of the graph. The paper
\cite{tegling2019scalability} considers \textbf{globally bounded neighborhoods}
in the sense that there exists a \textbf{global} bound $q$ for the in-degree:
\[
d_{\text{in}}(i) < q
\]
for all agents $i=1,\ldots, N$.

In scale-free protocols, we are looking for protocols which do not
depend on the network structure. This is motivated by the fact that in
many applications, an agent might be added/removed or a link might
fail and you then do not want to have to redesign the protocols being
used. This makes using such a \textbf{global} bound undesirable.

However, in most cases it turns out that it is sufficient if agent $i$
has a local bound available to $d_{in}(i)$.  Note that this is
actually a reasonable assumption because it is really a \textbf{local}
bound since it only bounds the weight of the edges going into node $i$
and does not rely on the rest of the network.

We refer to the property where agent $i$ has a bound $q_i$ available with
\begin{equation}\label{qibound}
	q_i > d_{\text{in}}(i)
\end{equation}
for $i=1,\ldots,N$ as \textbf{locally bounded neighborhoods}. In that
case, we can define
\[
\tilde{\zeta}_i(t) = \frac{1}{1+q_i} \zeta_i(t)
\]
and we obtain:
\begin{equation}\label{zeta-y}
	\tilde{\zeta}_i(t)=\sum_{j=1,j\neq i}^N d_{ij}(y_i(t)-y_j(t)),
\end{equation}
where
\[
d_{ij}= \frac{a_{ij}}{1+q_i},
\]
for $i\neq j$ while 
\[
d_{ii}=1-\sum_{j=1,j\neq i}^Nd_{ij}
\]
Note that the weight matrix $D=[d_{ij}]$ is then a, so-called, row
stochastic matrix. If we design a protocol based on these scaled data:
\begin{equation}\label{protoco2}
	\begin{system}{cl}
		x_{i,c}^+ &=A_{c}x_{i,c}+B_{c}\tilde{\zeta}_i,\\
		u_i &=F_{c}x_{i,c},
	\end{system}
\end{equation}
where $A_c$, $B_c$ and $F_c$ are independent of the network structure
then we implement this protocol for the original network as:
\begin{equation}\label{protoco3}
	\begin{system}{cl}
		x_{i,c}^+ &=A_{c}x_{i,c}+\frac{1}{1+q_i} B_{c}\zeta_i,\\
		u_i &=F_{c}x_{i,c},
	\end{system}
\end{equation}
where $x_{c,i}(t)\in\R^{n_c}$ is the state of protocol.

Traditionally, in continuous-time multi-agent systems we have used the
Laplacian matrix and we have not used these local bounds. On the other
hand for discrete-time multi-agent systems in the literature we have
always used the row stochastic matrix and we therefore implicitly
assumed knowledge of these local bounds on the network.

This paper will investigate, for both continuous-time and discrete-time
multi-agent systems, whether the use of these local bounds can improve
design possibilities for scale-free protocols that achieve synchronization.

We first need a definition before we give a precise problem
formulation.

\begin{definition}\label{def1}  
	We define the set $\mathbb{G}^N$ as the set of all
	directed graphs of $N$ agents which contain a directed
	spanning tree.
\end{definition}

We formulate the scale-free or scale-free synchronization problem of a
MAS as follows.

\begin{problem}\label{prob4}
	The \textbf{scale-free state synchronization problem without local
		bounds} for MAS \eqref{eq1} with communication given by
	\eqref{zeta} is to find, if possible, a fixed linear protocol of the
	form:
	\begin{equation}\label{protoco1}
		\begin{system}{cl}
			x_{i,c}^+ &=A_{c}x_{i,c}+B_{c}\zeta_i,\\
			u_i &=F_{c}x_{i,c},
		\end{system}
	\end{equation}
	where  $x_{c,i}(t)\in\R^{n_c}$ is the state of protocol, such that
	state synchronization is achieved, i.e.
	\begin{equation}\label{synch_state}
		\lim_{t\rightarrow \infty} x_i(t)-x_j(t) =0
	\end{equation}
	for all $i,j=1,\ldots,N$ for any number of agents $N$, for
	any fixed communication graph $\mathcal{G}\in\mathbb{G}^N$ and for
	all initial conditions of agents and protocols.
\end{problem}

\begin{problem}\label{prob5}
	The \textbf{scale-free state synchronization problem with local
		bounds} for MAS \eqref{eq1} with communication given by
	\eqref{zeta} is to find, if possible, a fixed linear protocol of the form
	\eqref{protoco3} , such that state synchronization
	\eqref{synch_state} is achieved for any number of agents $N$, any
	$q_1,\ldots q_N \in \R^+$, for any fixed communication graph
	$\mathcal{G}\in\mathbb{G}^N$ satisfying \eqref{qibound} and for all
	initial conditions of agents and protocols.
\end{problem}

In both problems the protocol parameters $A_c$, $B_c$ and $F_c$ are
designed completely independent of the network structure. They only
rely on the agent model, i.e. $A$, $B$ and $C$.  The only but
intrinsic difference between these two problems is that in Problem
\ref{prob5} we added an initial scaling of the measurements based on a
local bound for the weighted in-degree.

Effectively, in Problem \ref{prob4} we use communication described by
a Laplacian matrix while in Problem \ref{prob5} by using
\eqref{zeta-y} we use communication described by a row-stochastic
matrix which requires the availability of these local
bounds. Classically, Problem \ref{prob4} would be standard for
continuous-time systems and Problem \ref{prob5} would be standard for
discrete-time systems. We should note that in many papers discrete-time
problems are immediately defined in terms of the row-stochastic matrix
without making the scaling explicit.

\section{Continuous-time results}

As indicated before we are going to investigate solvability of
problems \ref{prob4} and \ref{prob5} for continuous-time systems. We
will in the next subsection consider Problem \ref{prob4} where we use
the classical Laplacian matrix and then we will consider in the
subsection thereafter Problem \ref{prob5} where we used the local
bounds to convert the Laplacian matrix into a row-stochastic matrix.

\subsection{Scale-free synchronization without locally bounded neighborhoods}

In this section, we consider scale-free synchronization for
continuous-time systems with a network described by a Laplacian
matrix. We want to find conditions whether Problem \ref{prob4} is
solvable for this class of systems. In other words, we have no
information available about the network and want to design a protocol
that achieves synchronization. We first establish necessary conditions
and then sufficient conditions.

\subsubsection{Necessary conditions}

\begin{theorem}\label{theorem1}
  Consider the scale-free state synchronization problem without local
  bounds as formulated in Problem \ref{prob4} for continuous-time systems.
  \begin{description}
  \item[Part 1:] The scale-free state synchronization problem without
    local bounds is solvable with a scalar input and/or a scalar
    output \emph{only if} the agent model \eqref{eq1} is either
    asymptotically stable or satisfies the following conditions:
    \begin{enumerate}
    \item Stabilizable and detectable,
    \item Neutrally stable,
    \item Weakly minimum phase,
    \item Relative degree equal to 1.
    \end{enumerate}
  \item[Part 2:] The scale-free state synchronization problem without
    local bounds is solvable for multi-input/multi-output agents
    \emph{only if} the agent model \eqref{eq1} is:
    \begin{enumerate}
    \item Stabilizable and detectable,
    \item All poles are in the closed left half plane.
    \end{enumerate}
  \end{description}
\end{theorem}

\begin{remark}
  The paper \cite[Theorem 4.1]{tegling-bamieh-sandberg-auto-23} shows
  that there is no linear \textbf{static} protocol which can achieve
  scale-free synchronization for MAS modeled as a chain of $n$
  integrators ($n\geq 2$) with full state coupling. Theorem
  \ref{theorem1} in this paper extends the result of
  \cite{tegling-bamieh-sandberg-auto-23}. We prove that there is no
  linear \textbf{dynamic} protocol either to achieve scale-free
  synchronization for this class of MAS.
\end{remark}

	\begin{remark}
	In \cite{liu-saberi-stoorvogel-tac-2023}, we provided necessary
	conditions for MAS consisting of SISO agent. Here we have obtained
	the necessary conditions for MAS consisting of SIMO, MISO, or MIMO
	agents. 
\end{remark}
	
\begin{remark} We would like to point out that the conditions of
  neutral stability and weakly minimum phase, which are necessary in
  the SISO case, are not necessary in the MIMO case as illustrated in
  following examples.
\end{remark}

\begin{example}\label{example1}
	To illustrate that neutrally stable is not a necessary condition
	for MIMO systems consider the system \eqref{eq1} with:
	\[
	A=\begin{pmatrix} 0 & 1 \\ 0 & 0 \end{pmatrix},\quad
	B=\begin{pmatrix} 1 & 0 \\ 0 & 1 \end{pmatrix},\quad
	C=\begin{pmatrix} 1 & 0 \\ 0 & 1 \end{pmatrix}.
	\]
	Clearly, the system is not neutrally stable but it is easy to verify
	that the protocol $u_i=-\zeta_i$ will achieve scale-free state
	synchronization.
\end{example}      

\begin{example}\label{example2}
	To illustrate that in the MIMO case in some peculiar cases the
	system can even have zeros in the open right half plane while
	scale-free state synchronization problem is still possible, consider
	the system \eqref{eq1} with:
	\[
	A=\begin{pmatrix} 0 & 0 & 1 \\ 0 & -1 & 1 \\ 0 & 0 &
		-1  \end{pmatrix},\quad 
	B=\begin{pmatrix} 1 & 0 \\ 0 & 0 \\ 0 & 1  \end{pmatrix},\quad
	C=\begin{pmatrix} 1 & 0 & 0 \\ 0 & 1 & -2 \end{pmatrix}.
	\]
	It is easily verified that the system has an invariant zero in $1$
	but the protocol
	\[
	u_i=\begin{pmatrix} -1 & 0 \\ 0 & 0 \end{pmatrix}\zeta_i
	\]
	will achieve scale-free state synchronization. Effectively, we see
	that we can have non-minimum phase agents, if the agent with
	transfer matrix $G$ can be stabilized by a
	controller with transfer matrix $G_c$ such that $GG_c$ has no unstable
	zeros. In other words, the unstable zero can be canceled without an
	unstable pole-zero cancellation. This only happens in cases such as
	the above example where one channel contains stable, non-minimum
	phase dynamics and another channel contains unstable, minimum
	phase dynamics. Clearly this cannot be done in a MISO, SIMO or SISO
	system where we effectively have only one channel available for
	feedback.
\end{example}      

\begin{proof}[Proof of Theorem \ref{theorem1}]
	The necessity of stabilizability and detectability
	is obvious. By using protocol \eqref{protoco1} and defining
	\begin{equation}\label{ABCtilde}
		\tilde{A}=\begin{pmatrix} A & BF_c \\ 0 & A_c \end{pmatrix},\qquad
		\tilde{B}=\begin{pmatrix} 0 \\ B_c \end{pmatrix},\qquad
		\tilde{C}=\begin{pmatrix} C & 0 \end{pmatrix}
	\end{equation}
	then \cite[Chapter 3]{saberi-stoorvogel-zhang-sannuti} has shown
	that we achieve synchronization if
	\[
	\tilde{A}+\lambda_i \tilde{B}\tilde{C}
	\]
	is Hurwitz stable for all nonzero eigenvalues
	$\{\lambda_2,\ldots,\lambda_N\}$ of the Laplacian matrix $L$. Since,
	we are looking for a scale-free protocol which works for any network
	in $\mathbb{G}^N$. we need that
	\begin{equation}\label{abc}
		\tilde{A}+\lambda \tilde{B}\tilde{C}
	\end{equation}
	is Hurwitz stable for all $\lambda\in \C$ with
	$\re \lambda >0$.
	
	The SISO result has been presented before in
	\cite[Theorem 1]{liu-saberi-stoorvogel-tac-2023}. In the
	multi-input and single-output case, we can follow the arguments
	provided in the proof of that paper to conclude that,
	for an agent with transfer matrix $G$ /and a protocol with transfer
	agent $G_c$, we need that $GG_c$ (which is a scalar rational
	function) is positive-real and hence needs to be neutrally
	stable, weakly minimum-phase and relative degree $1$.
	
	Clearly the Hurwitz stability of \eqref{abc} requires the transfer
	matrix of the system:
	\[
	\dot{p}=(\tilde{A}+\lambda \tilde{B}\tilde{C})p+\begin{pmatrix} B
		\\ 0 \end{pmatrix} v,\qquad\quad
	z= \tilde{C}p
	\]
	to be asymptotically stable which implies:
	\begin{equation}\label{GGc1}
		(I-\lambda GG_c)^{-1}G
	\end{equation}
	is asymptotically stable. If $G$ has a repeated
	pole on the imaginary axis then this can only be cancelled by
	the scalar transfer function $(I-\lambda GGc)^{-1}$ if $GG_c$ has a
	repeated pole on the imaginary axis which leads to a
	contradiction since $GG_c$ was neutrally stable.
	
	Similarly the Hurwitz stability of \eqref{abc} requires the transfer
	matrix of the system:
	\[
	\dot{p}=(\tilde{A}+\lambda \tilde{B}\tilde{C})p+\tilde{B}v,\qquad\quad
	z= \begin{pmatrix} 0 & F_c \end{pmatrix} p
	\]
	to be asymptotically stable which implies:
	\begin{equation}\label{GGc2}
		G_c(I-\lambda GG_c)^{-1}
	\end{equation}
	is asymptotically stable and strictly proper.
	
	If $G$ has a repeated invariant zero $s_0$ on the imaginary axis
	then for $GG_c$ to be weakly minimum-phase we need that $G_c$ has a
	pole in $s_0$. It can be easily verified that this yields a
	contradiction with \eqref{GGc2} being asymptotically stable.
	
	Finally, if $G$ has relative degree $2$ or higher, then $GG_c$ can
	never have relative degree $1$ for a strictly proper protocol of the
	form \ref{protoco1}.
	
	The above argument can be easily modified for the
	single-input and multi-output case, where we again follow the arguments
	provided in the proof of \cite[Theorem
	1]{liu-saberi-stoorvogel-tac-2023} to conclude this time that $G_cG$
	(instead of $GG_c$) is positive-real and hence needs to be neutrally
	stable, weakly minimum-phase and relative degree $1$. Since, in this
	case, $G_cG$ is a scalar transfer function the rest of the above
	arguments can be easily modified.
	
	For MIMO systems, we also need that \eqref{abc}
	is Hurwitz stable for all $\lambda\in \C$ with
	$\re \lambda >0$. Since $\lambda$ can be arbitrarily small it is
	obvious that \eqref{abc} Hurwitz stable for all
	$\lambda\in \C$ with $\re \lambda >0$ requires that the eigenvalues
	of $\tilde{A}$ have to be in the closed left half plane which
	trivially implies that the eigenvalues of $A$ have to be in
	the closed left half plane
\end{proof}

\subsubsection{Sufficient conditions}

\begin{theorem}\label{theorem2}
	The scale-free state synchronization problem without local bounds as
	formulated in Problem \ref{prob4} is solvable if the continuous-time
	agent model \eqref{eq1} is either asymptotically stable or satisfies
	the following conditions:
	\begin{enumerate}
		\item Stabilizable and detectable,
		\item Neutrally stable,
		\item Minimum phase,
		\item Agent model has uniform rank with order of infinite zero
		equal to one.
	\end{enumerate}
\end{theorem}

\begin{remark}
	Note that the sufficient conditions of Theorem \ref{theorem2} are
	very close to the necessary conditions of Theorem \ref{theorem1} for
	single-input or single-output systems. We only strengthen the
	requirement of weakly minimum phase to minimum phase.
	
	For MIMO systems the gap between necessary and sufficient conditions
	is much larger but only because of some very peculiar cases like the
	ones in Example \ref{example1} and \ref{example2}. We claim that
	generically the necessary conditions for the SISO case also apply in
	the MIMO case.
\end{remark}

\begin{proof}
	This result has been presented before in \cite[Theorem
	3]{liu-saberi-stoorvogel-tac-2023}.
\end{proof}

Note that in \cite{liu-saberi-stoorvogel-tac-2023} we have also
presented simulations to see how these protocols perform.

\subsection{Scale-free synchronization with locally bounded neighborhoods}

\subsubsection{Necessary conditions}

\begin{theorem}\label{theorem3}~
  Consider the scale-free state synchronization problem with local
  bounds as formulated in Problem \ref{prob5}  for
  continuous-time agents.
  \begin{description}
  \item[Part 1:] The scale-free state synchronization problem with
    local bounds is solvable with a scalar input and/or a scalar output
    \emph{only if} the agent model \eqref{eq1} is either
    asymptotically stable or satisfies the following conditions:
    \begin{enumerate}
    \item Stabilizable and detectable,
    \item Neutrally stable.
    \end{enumerate}
  \item[Part 2:] The scale-free state synchronization
    problem with local bounds is solvable for MIMO agents
    \emph{only if} the agent model \eqref{eq1} is:
    \begin{enumerate}
    \item Stabilizable and detectable,
    \item All poles are in the closed left half plane.
    \end{enumerate}
  \end{description}
\end{theorem}

\begin{remark}
	For agents with a scalar input and/or a scalar output the conditions
	are actually necessary and sufficient as we will see in the next
	subsection.
\end{remark}

\begin{proof}
	The necessity of stabilizability and detectability is obvious. By
	using protocol \eqref{protoco2} and defining \eqref{ABCtilde} then
	\cite[Chapter 3]{saberi-stoorvogel-zhang-sannuti} has shown that we
	achieve synchronization if
	\begin{equation}\label{abc3}
		\tilde{A}+(1-\lambda_i) \tilde{B}\tilde{C}
	\end{equation}
	is Hurwitz stable for all eigenvalues
	$\{\lambda_2,\ldots,\lambda_N\}$ of the row
	stochastic matrix $D$ unequal to $1$. For a network in $\mathbb{G}^N$, we find that
	$|\lambda_i|<1$ for $i=2,\ldots, N$. Moreover, it is clear that for
	any $\lambda$ with $|\lambda| <1$, there exists a network in
	$\mathbb{G}^N$ whose associated row stochastic matrix has an
	eigenvalue in $\lambda$.
	
	Since, we are looking for a scale-free protocol which works for any
	network in $\mathbb{G}^N$. we need that
	\begin{equation}\label{abc4}
		\tilde{A}+(1-\lambda) \tilde{B}\tilde{C}
	\end{equation}
	is Hurwitz stable for all $\lambda$ with $|\lambda|<1$.
	
	Using arguments similar to \cite[Theorem
	2]{liu-saberi-stoorvogel-tac-2023} we note that we need that
	$GG_c+\tfrac{1}{2}$ is positive real (single-output case) or
	$G_cG+\tfrac{1}{2}$ is positive real (single-input case). We can
	then use similar arguments as in the proof of Theorem \ref{theorem1}
	to conclude that $G$ needs to be neutrally stable.

	For the general MIMO case we note that in \eqref{abc4}, the
	$\lambda$ can be arbitrarily close to $1$, and therefore the
	eigenvalues of $\tilde{A}$ have to be in the closed left half plane
	which trivially implies that the eigenvalues of $A$ have to be in
	the closed left half plane
\end{proof}

\subsubsection{Sufficient  conditions}


\begin{theorem}\label{theorem4}
	The scale-free state synchronization problem with local bounds as
	formulated in Problem \ref{prob5} is solvable if the continuous-time
	agent model \eqref{eq1} is either asymptotically stable or satisfies
	the following conditions:
	\begin{enumerate}
		\item Stabilizable and detectable,
		\item Neutrally stable.
	\end{enumerate}
\end{theorem}

\begin{remark}
	We see that the availability of these local bounds and being able to
	convert the Laplacian matrix into a row-stochastic matrix allows us
	to solve our problem without imposing any constraints on the
	invariant zeros of the system or the relative degree.
	
	The only gap between our necessary and sufficient conditions is for
	MIMO systems where necessity only requires the poles to be in the
	closed left half plane while our sufficient conditions impose
	neutral stability. That neutral stability is not necessary in this
	case is illustrated by the system in Example \ref{example1}.
\end{remark}

\begin{proof}
	We first note that since the system is neutrally stable there
	exists $P>0$ such that
	\[
	A\T P +PA \leq 0.
	\]
	Next, we consider the protocol
	\begin{equation}\label{protocol2}
		\begin{system*}{ccl}
			\dot\chi_i &=& (A+HC)\chi_i-H\tilde{\zeta}_i \\
			u_i &=& -\delta B\T P \chi_i
		\end{system*}
	\end{equation}
	where $H$ is such that $A+HC$ is Hurwitz stable and $\delta>0$
	needs to be small enough as will become clear later.
	
	Since $A+HC$ is Hurwitz stable there exists $\eps>0$ and $Q>0$
	such that
	\begin{equation}\label{lyap}
		(A+HC)\T Q+Q(A+HC)+\eps Q +I =0
	\end{equation}
	Next we choose $\delta>0$ such that:
	\begin{equation}\label{boun}
		2\delta (QBB\T P + PBB\T Q) \leq \eps Q
	\end{equation}
	By using protocol \eqref{protocol2} and defining
	\begin{equation}\label{ABCtilde2}
		\tilde{A}=\begin{pmatrix} A & -\delta BB\T P \\
			0 & A+HC \end{pmatrix},\qquad
		\tilde{B}=\begin{pmatrix} 0 \\ -H \end{pmatrix},\qquad
		\tilde{C}=\begin{pmatrix} C & 0 \end{pmatrix}
	\end{equation}
	then (as argued in the proof of Theorem \ref{theorem3}), we
	need that
	\begin{equation}\label{abc2}
		\tilde{A}+(1-\lambda) \tilde{B}\tilde{C}
	\end{equation}
	is Hurwitz stable for all $\lambda$ with $|\lambda|<1$.
	
	We need to prove that the interconnection of \eqref{protocol2} and
	\eqref{eq1} with $\tilde{\zeta}_i=(1-\lambda)y_i$ is asymptotically
	stable for all $\lambda$ with $|\lambda|<1$. The dynamics
	associated with the matrix \eqref{abc2} are given by:
	\[
	\begin{system*}{ccl}
		\dot{\varphi} &=&  A\varphi -\delta BB\T P \psi\\
		\dot{\psi} &=& (A+HC)\psi -(1-\lambda)HC\varphi 
	\end{system*}
	\]
	Choosing
	\[
	\tilde{\varphi}=(1-\lambda)\varphi\quad \text{ and }\quad
	e=\psi-\tilde{\varphi}
	\]
	we obtain:
	\[
	\begin{system*}{ccl}
		\dot{\tilde{\varphi}} &=& A\tilde{\varphi} -\delta (1-\lambda)
		BB\T P (e+\tilde{\varphi}) \\
		\dot{e} &=& (A+HC)e + \delta (1-\lambda)
		BB\T P (e+\tilde{\varphi})
	\end{system*}
	\]
	We note that \eqref{lyap} and \eqref{boun} imply that:
	\[
	[A+HC+\delta(1-\lambda)BB\T P]^* Q + Q [A+HC+\delta(1-\lambda)BB\T
	P] + I \leq 0 
	\]
	Define $V_1=e^*Qe$ and we obtain:
	\begin{align*}
		\dot{V}_1 &\leq -e^* e + \delta (1-\lambda) e^* QBB\T P\tilde{\varphi} +
		\delta (1-\lambda^*) \tilde{\varphi}^* PBB\T Qe\\
		&\leq -e^* e + 4\delta e^* QBB\T Q e +
		\tfrac{1}{4} \delta | 1-\lambda|^2 \tilde{\varphi}^* PBB\T P
		\tilde{\varphi} 
	\end{align*}
	Next, we consider $V_2=\tilde{\varphi}^*P\tilde{\varphi}$ and we
	obtain:
	\begin{align*}
		\dot{V}_2 &= \tilde{\varphi}^*(A\T P+PA)\tilde{\varphi}-\delta
		(1-\lambda) 
		\tilde{\varphi}^* PBB\T P (e+\tilde{\varphi})
		-\delta (1-\lambda^*)(e+\tilde{\varphi})^* PBB\T P \tilde{\varphi}\\
		&\leq -\delta | 1-\lambda|^2 \tilde{\varphi}^* PBB\T P
		\tilde{\varphi} -\delta (1-\lambda)
		\delta \tilde{\varphi}^* PBB\T P e
		-\delta (1-\lambda^*)e^* PBB\T P \tilde{\varphi}\\
		&\leq -\tfrac{3}{4} \delta | 1-\lambda|^2 \tilde{\varphi}^* PBB\T P
		\tilde{\varphi} + 4 \delta e^* PBB\T P e  
	\end{align*}
	where we used that $|\lambda|<1$ implies
	\begin{equation}\label{dfbound}
		|1-\lambda|^2\leq 2\re(1-\lambda).
	\end{equation}
	Combining the above inequalities and choose$\delta$ small
        enough such that $8 \delta (PBB\T P+QBB\T Q) < I$, we find:
	\[
	\dot{V}_1+\dot{V}_2 \leq -\tfrac{1}{2} e^* e - \tfrac{1}{2}
	\delta | 1-\lambda|^2 \tilde{\varphi}^* PBB\T P 
	\tilde{\varphi}
	\]
	from which asymptotically stability follows using a standard
	argument based on LaSalle invariance principle.
\end{proof}                     

\section{Discrete--time results}

Next, we are going to investigate solvability of problems \ref{prob4}
and \ref{prob5} for discrete-time systems. We will in the next
subsection consider Problem \ref{prob4} where we use the Laplacian
matrix and then we will consider in the subsection thereafter Problem
\ref{prob5} where we used the local bounds to convert the Laplacian
matrix into a row-stochastic matrix. The latter is the classical case
for discrete-time systems.

\subsection{Scale-free synchronization without locally bounded neighborhoods}

\begin{theorem}\label{theorem5}
	The scale-free state synchronization problem without local bounds as
	formulated in Problem \ref{prob4} is NOT solvable except for the
	trivial case when the agents are asymptotically stable.
\end{theorem}

\begin{proof}
	Consider a protocol \eqref{protoco1}. Then similarly as in the proof
	of Theorem \ref{theorem1}, we can define \eqref{ABCtilde} and argue that
	\[
	\tilde{A}+\lambda \tilde{B}\tilde{C}
	\]
	must be Schur stable (eigenvalues in open unit disc) for
	all $\lambda\in \C$ with $\re \lambda >0$. Assume this is
	true. Consider:
	\begin{equation}\label{rrtY}
		\det(sI-\tilde{A}-\lambda \tilde{B}\tilde{C})
	\end{equation}
	If this determinant does not depend on $\lambda$ then
	\[
	\det(sI-\tilde{A}+\lambda \tilde{B}\tilde{C})=\det(sI-\tilde{A})
	\]
	and if this is asymptotically stable then $\tilde{A}$ must be
	Schur stable which is only possible if the matrix $A$ is
	already Schur stable. Note that the coefficients of a
	characteristic polynomial of a Schur stable matrix of fixed
	dimensions can never exceed a certain bound $M$ since all
	eigenvalues are bounded. But if \eqref{rrtY} depends on $\lambda$
	then for large enough $\lambda$ the coefficients of this
	characteristic polynomial will exceed this bound $M$ which implies
	that the matrix is not Schur stable which yields a contradiction.
\end{proof}

\subsection{Scale-free synchronization with locally bounded neighborhoods}

\subsubsection{Necessary conditions}

\begin{theorem}\label{theorem6}
  Consider the scale-free state synchronization problem with local
  bounds as formulated in Problem \ref{prob5} for discrete-time
  systems.
  \begin{description}
  \item[Part 1:] The scale-free state synchronization problem with
    local bounds is solvable for single-input or single-output agents
    \emph{only if} the agent model \eqref{eq1} is either
    asymptotically stable or satisfies the following conditions:
    \begin{enumerate}
    \item Stabilizable and detectable,
    \item Neutrally stable.
    \end{enumerate}
  \item[Part 2:] The scale-free state synchronization problem with
    local bounds is solvable for MIMO agents \emph{only if} the agent
    model \eqref{eq1} is:
    \begin{enumerate}
    \item Stabilizable and detectable,
    \item All poles are in the closed unit disc.
    \end{enumerate}
  \end{description}
\end{theorem}

\begin{remark}
	In the single-input or single-output case the conditions are
	actually necessary and sufficient as we will see in the next
	subsection.
\end{remark}

\begin{proof}
  The arguments are identical to the proof of the continuous-time
  result in Theorem \ref{theorem3}. In other words, we achieve
  synchronization if
  \begin{equation*}
    \tilde{A}+(1-\lambda_i) \tilde{B}\tilde{C}
  \end{equation*}
  is Schur stable for all eigenvalues
  $\{\lambda_2,\ldots,\lambda_N\}$ unequal to $1$ of the row
  stochastic matrix $D$.
	
  Using arguments similar to \cite[Theorem
  2]{liu-saberi-stoorvogel-tac-2023} we note that we need that
  $GG_c+\tfrac{1}{2}$ is positive real (single-output case) or
  $G_cG+\tfrac{1}{2}$ is positive real (single-input case). We can
  conclude that $G$ needs to be neutrally stable.
	
  For the general MIMO case we note that the $\lambda$ can be
  arbitrarily close to $1$, and therefore the eigenvalues of
  $\tilde{A}$ have to be in the closed unit disc which trivially
  implies that the eigenvalues of $A$ have to be in the closed unit
  disc.
\end{proof}

\subsubsection{Sufficient conditions}

\begin{theorem}\label{theorem7}
	The scale-free state synchronization problem with local bounds as
	formulated in Problem \ref{prob5} is solvable if the discrete-time
	agent model \eqref{eq1} is either asymptotically stable or satisfies
	the following conditions:
	\begin{enumerate}
		\item Stabilizable and detectable,
		\item Neutrally stable.
	\end{enumerate}
\end{theorem}

\begin{proof}
	This result has already been presented in \cite[Theorem
	5]{liu-saberi-stoorvogel-tac-2023}.
\end{proof}

\section{Numerical examples}

We have three theorems in this paper dealing with sufficient
conditions. For Theorems 2 and 7 there are already illustrative
numerical examples for the design in
\cite{liu-saberi-stoorvogel-tac-2023}. In this section, we choose a
SISO agent model that is nonminimum phase with relative degree higher
than one. As such, based on Theorem \ref{theorem1}, scale-free
protocol can not be designed without the knowledge of local
bounds. However in the following example, we investigate the protocol
design given in the proof of Theorem \ref{theorem4} and illustrate the
results for two different communication graphs.

\subsection{Agent model}
Consider a continuous-time MAS with SISO agent model $(C, A, B)$:
\begin{equation*}
	A=\begin{pmatrix}
		0&1&1\\-1&0&1\\0&0&0
	\end{pmatrix},\  B=\begin{pmatrix}
	0\\2\\-1
	\end{pmatrix},\ C=\begin{pmatrix}
		1&0&0
	\end{pmatrix}.
\end{equation*}
Obviously, this model is stablizable and detectable, and neutrally
stable. On the other hand, it is non-minimum phase and has relative
degree 2. Obviously, scale-free protocol can not be designed based on
Theorem \ref{theorem2}. Thus, we use protocol \eqref{protocol2} to
obtain the corresponding result.

\subsection{Protocol design}

We design the protocol shown in \eqref{protocol2} as follows,
\[
\begin{system}{ccl}
	\dot\chi_i &=&\begin{pmatrix}
		-1&1&1\\-1&0&1\\-1&0&0
	\end{pmatrix}\chi_i+\frac{1}{1+q_i}\begin{pmatrix}
		1\\0\\1
	\end{pmatrix}{\zeta}_i \\
	u_i &=& -\delta \begin{pmatrix}
		1&1&-1
	\end{pmatrix} \chi_i
\end{system}
\]
with $\delta=0.1$ and
\begin{equation*}
	P=\begin{pmatrix}
		1&0 &-1 \\
		0& 1& 1\\
		-1& 1& 3
	\end{pmatrix},
\end{equation*}
where the local bounds $q_i$ are chosen based on the adjacency matrix
of communication network.

\subsection{Two communication networks}

We consider two communication networks with different topologies to
show the efficacy of our protocols.

\begin{figure}[ht]
	\includegraphics[width=6.5cm, height=1.cm]{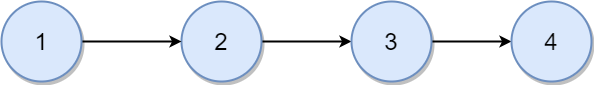}
	\centering
	\caption{Directed topology network with $4$ nodes}\label{graph1}
\end{figure}

\textbf{Case I:} We consider a MAS with $4$ agents (i.e.\ $N=4$), and
a directed communication topology shown in Figure \ref{graph1}.
If we assume that weighting values of adjacency matrix are all 1, i.e.,
\[
\mathcal{A}_{I}=\begin{pmatrix}
		0&0&0&0\\
		1&0&0&0\\
		0&1&0&0\\
		0&0&1&0
	\end{pmatrix}
\]
then we can choose $q_i=2$ for $i=1,2,3,4$.

\begin{figure}[ht]
	\includegraphics[width=6.5cm]{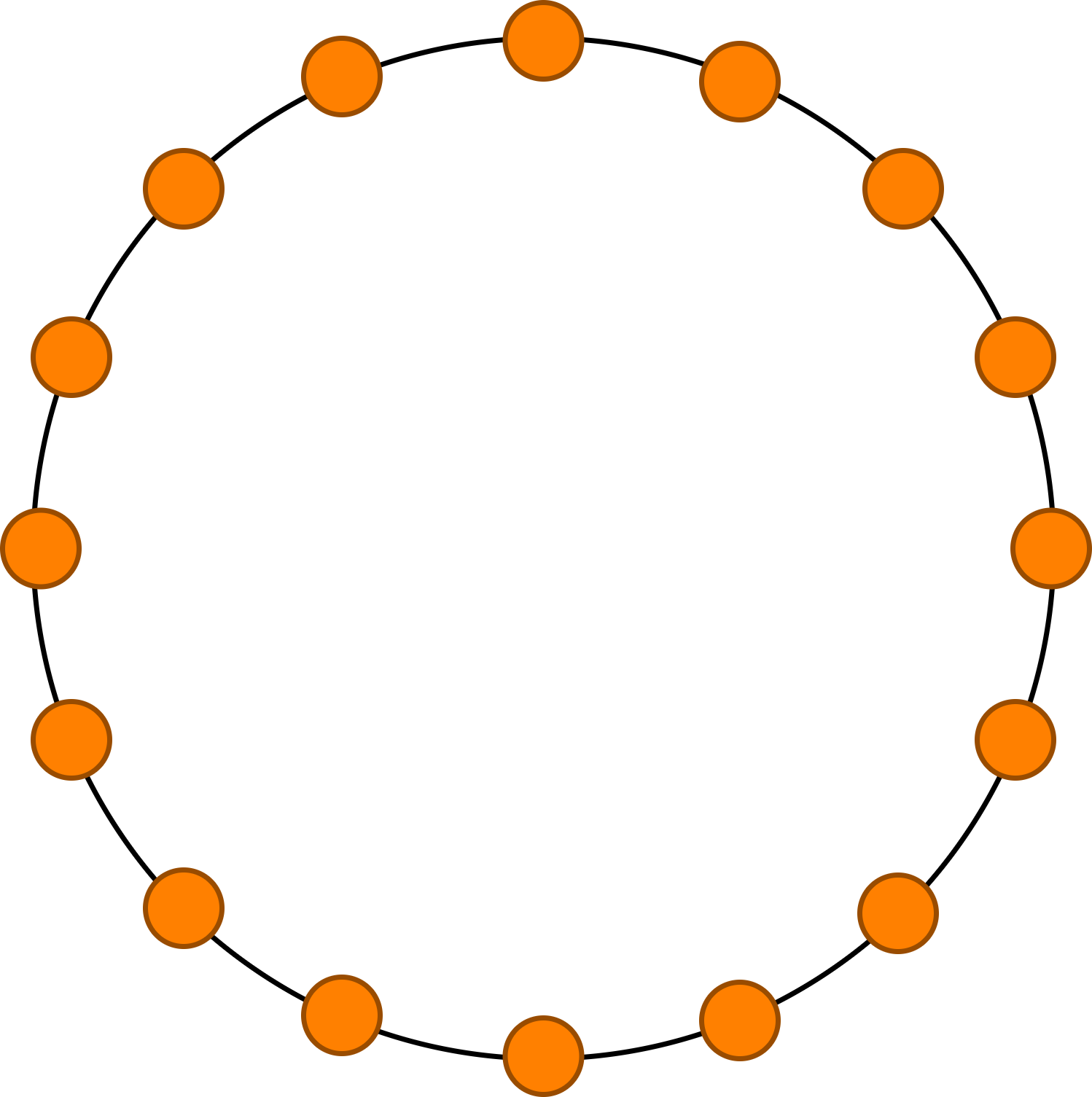}
	\centering
	\caption{Directed loop topology network with $60$ nodes}\label{graph2}
\end{figure}

\textbf{Case II:} In this case, we consider a MAS with $60$ agents
(i.e.\ $N = 60$) and a directed loop communication topology shown in Figure \ref{graph2}. 
When we have the associated adjacency matrix $\mathcal{A}_{II}$ with
$a_{i+1,i}=a_{1,60}=1$ and $i=1,\cdots,59$, then one can also choose $q_i=2$.

\subsection{Synchronization results}

\begin{figure}[ht!]
  \includegraphics[width=9.5cm]{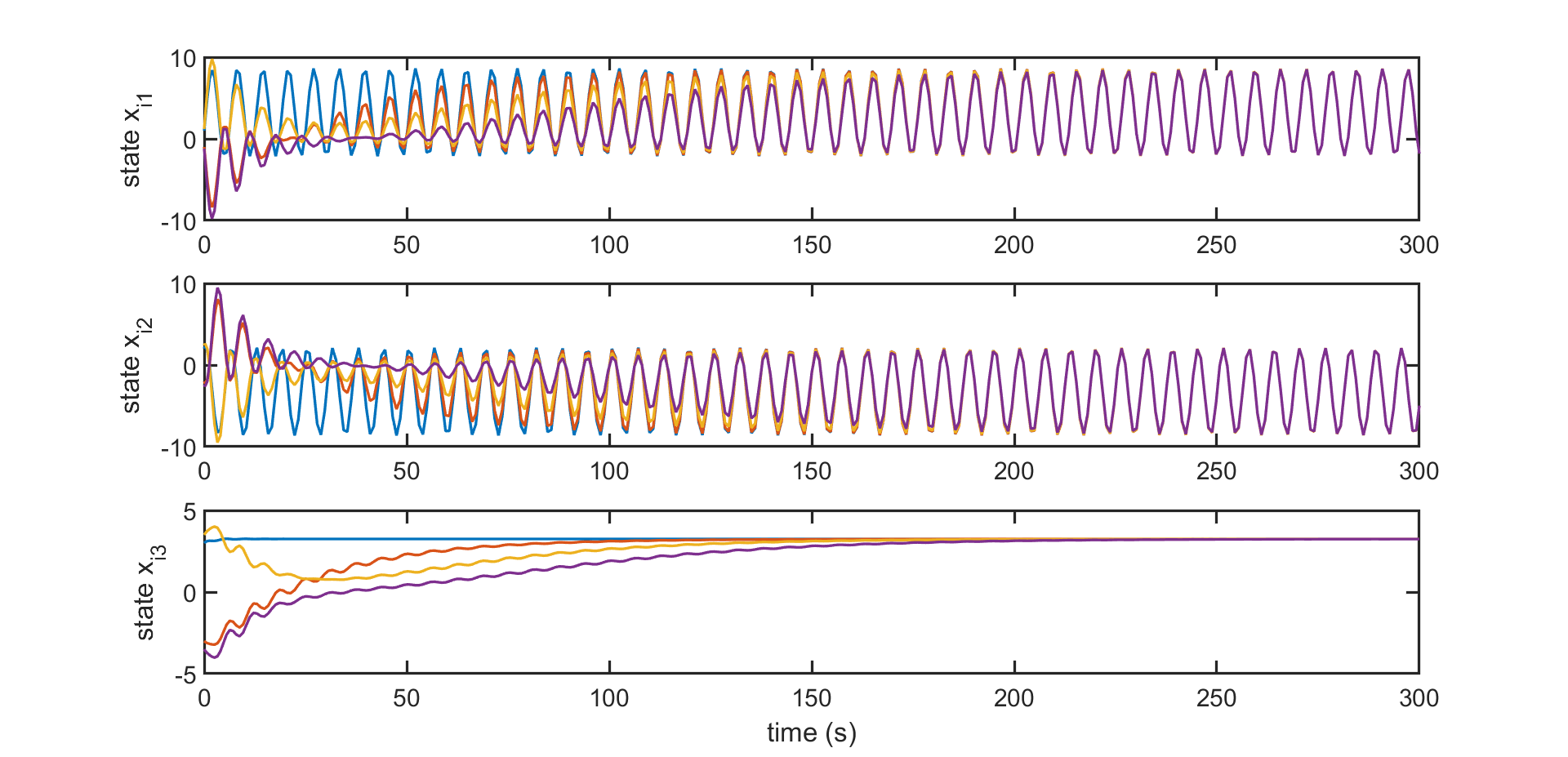}
  \centering
  \caption{State synchronization for continuous-time MAS with local
    bounded communication graph in Case I.}\label{con-case1}
\end{figure}

\begin{figure}[ht!]
  \includegraphics[width=9.5cm]{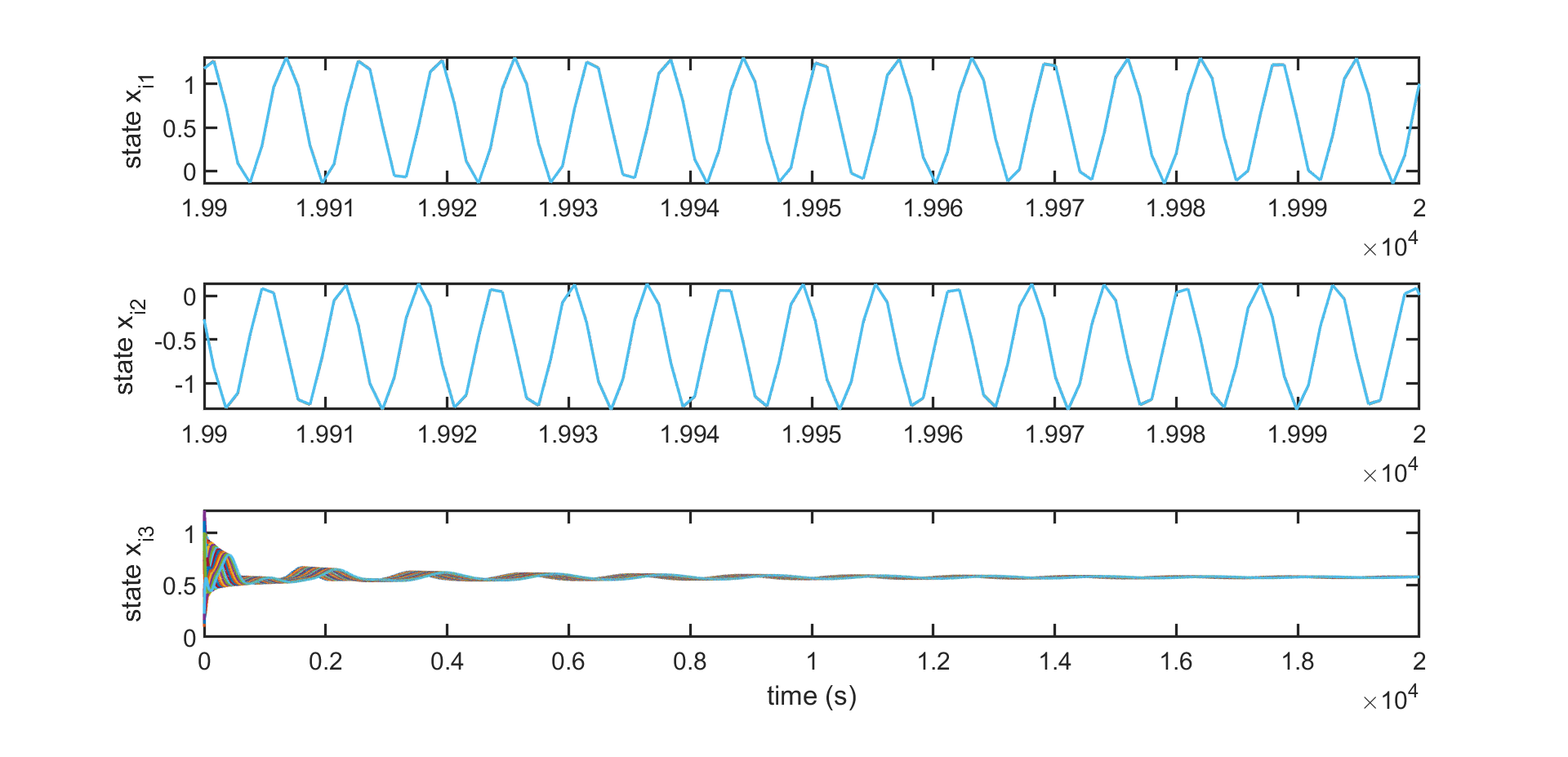} \centering
  \caption{State synchronization for continuous-time MAS with local
    bounded communication graph in Case II. For clarity of the graph,
    we only show the synchronized trajectories for states $x_{i1}$ and
    $x_{i2}$}\label{con-case2}
  \includegraphics[width=9.5cm]{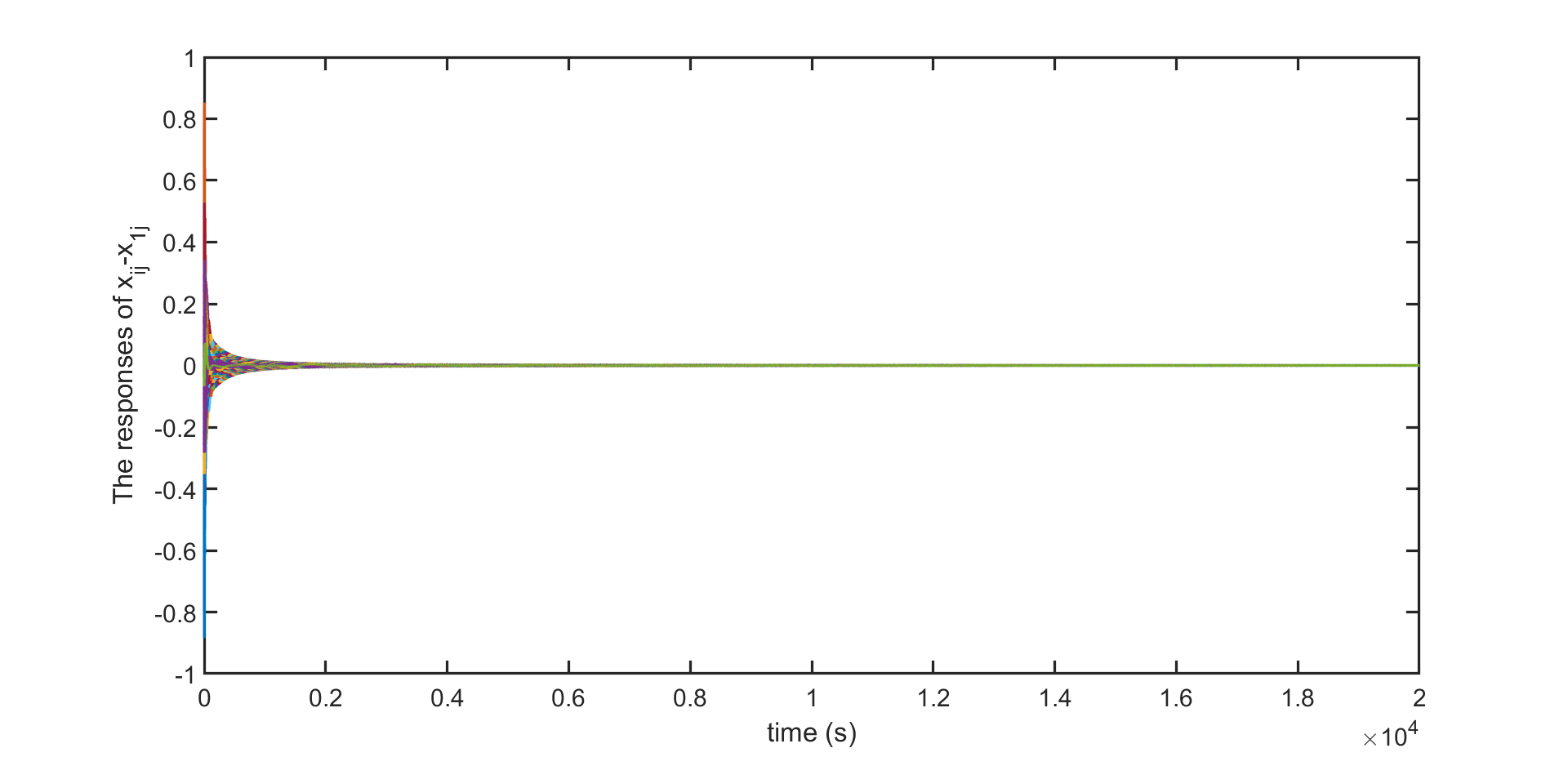} \centering
  \caption{Error between the states for continuous-time MAS with locally bounded
    communication graph in Case II.}\label{con-case2-e}
\end{figure}

The simulation results for both Cases I and II are demonstrated in
Fig. \ref{con-case1} and \ref{con-case2}. The error between the states
$x_{ij}-x_{i1}$ are shown in Fig. \ref{con-case2-e} to show the
synchronization more clearly. The results illustrate that the protocol
design is independent of the communication graph and is scale free so
that we can achieve synchronization with a one-shot protocol design, for
any graph with any number of agents.


\section{Conclusion}

In this paper we have provided necessary and sufficient conditions for
the existence of solutions to the state synchronization problem of
homogeneous MAS via scale-free linear dynamic non-collaborative
protocols without or with locally bounded neighborhoods for both
continuous- and discrete-time. The necessary and sufficient conditions
show that the scale-free state synchronization can be achieved for
this class of MAS.  On the one hand, for continue-time MAS, locally
bounded neighborhoods can relax the necessary conditions (i.e., weakly
minimum phase and relative degree one). However, there is no linear
dynamic design that can remove the condition of neutrally stable. On
the other hand, for discrete-time MAS, without locally bounded
neighborhood the scale-free design via linear protocols is not
possible. However, with locally bounded neighborhoods, we can achieve
scale-free synchronization for MAS with neutrally stable agents.

Finally, our result shows that scale-free design via a linear
dynamic non-collaborative protocol essentially requires agents to be
neutrally stable. However, the paper
\cite{liu-zhang-saberi-stoorvogel-ccdc-2018} shows a scale-free design
via nonlinear protocol is possible without the neutrally stable
condition in the case of full-state coupling. Our future research
focuses on obtaining necessary and sufficient conditions for
scale-free design via nonlinear protocols for the case of partial-state
coupling.

\bibliographystyle{plain}
\bibliography{referenc}
		
\end{document}